\documentclass{article}
\usepackage{ijcai23}

\usepackage{times}
\usepackage{soul}
\usepackage{url}
\usepackage[hidelinks]{hyperref}
\usepackage[utf8]{inputenc}
\usepackage[small]{caption}
\usepackage{graphicx}
\usepackage{amsmath, amssymb}
\usepackage{amsthm}
\usepackage{booktabs}
\usepackage{algorithm}
\usepackage{algpseudocode}
\usepackage[switch]{lineno}
\usepackage{comment}
\usepackage[table]{xcolor}

\usepackage{xcolor}

\title{Discrete Diffusion Probabilistic Models for Symbolic Music Generation}

\author{
Matthias Plasser$^1$
\and
Silvan Peter$^1$\And
Gerhard Widmer$^{1,2}$
\affiliations
$^1$Institute of Computational Perception, Johannes Kepler University Linz, Austria\\
$^2$LIT AI Lab, Linz Institute of Technology, Austria
\emails
\{firstname\}.\{lastname\}@jku.at
}

\begin{document}

\maketitle

\begin{abstract}

Denoising Diffusion Probabilistic Models (DDPMs) have made great strides in generating high-quality samples in both discrete and continuous domains \shortcite{DBLP:Dhariwal21,DBLP:Taylor22,DBLP:Mittal21}.
However, \textit{Discrete} DDPMs (D3PMs) have yet to be applied to the domain of Symbolic Music.
This work presents the direct generation of Polyphonic Symbolic Music using D3PMs \footnote{Our implementation, trained model weights and samples can be found at \url{https://github.com/plassma/symbolic-music-discrete-diffusion}}.
Our model exhibits state-of-the-art sample quality, according to current quantitative evaluation metrics, and allows for flexible infilling at the note level.
We further show, that our models are accessible to post-hoc classifier guidance, widening the scope of possible applications.
However, we also cast a critical view on
quantitative evaluation of music sample quality via statistical metrics, and present a simple algorithm that can confound our metrics with completely spurious, non-musical samples.

\end{abstract}

\section{Introduction}

Music generation denotes a distinct set of tasks in the burgeoning field of synthetic art generation.
It subsumes a wide range of approaches and problems, both artistic and technical.
Common modeling goals are creating wholly unconstrained musical material from a random seed, the sensible connection of existing music excerpts, or the continuation of a starting sequence.
In terms of content, music generation can aim at the creation of individual tracks of predefined instruments, full compositions, melody accompaniments, drum loops, and many other uses.


Perhaps surprisingly to practitioners of generative models not familiar with music,
the data domain of music generation is not inherent to the problem.
Researchers model music generation both in the symbolic -- our modeling
target -- and the audio domain.


Symbolic music representations build on discrete musical objects, such as notes, commonly encoded in MIDI or musicXML files.
The strengths of symbolic representations lie in their capacity to build on a wealth of thinking about musical objects and structure.
The resulting representations are lightweight; they encode musical pieces in hundreds to thousands of units.
However, symbolic representations do not directly encode sound.


On the other hand, the audio domain is the digital representation of sound.
Audio data consists of sequences of air pressure measurements sampled at isochronous intervals.
Its strengths lie in the near-complete representation of audible sound and the simple, unstructured format.
However, it's precisely this lack of structure and the length of typical sequences that pose difficult modeling problems.
Research in these two domains shares many similarities; however, they also have non-overlapping problems and diverging histories. 
A technical difference between the two music domains lies in their continuous versus discrete data distributions.
This difference is mirrored in the probabilistic formulation of many generative latent space models.
Among them are Diffusion Probabilistic Models (DPM) as introduced by  Sohl-Dickstein et al.~\shortcite{DBLP:SD15}.
Their most famous formulation --- denoising diffusion probabilistic models (DDPM) --- has been shown to outperform previous state-of-the-art generative models in various continuous generation tasks~\cite{DBLP:Dhariwal21,DBLP:Taylor22,DBLP:SD15,DBLP:Ho20}.
Consequentially, audio generation research applied DDPMs to the audio domains of speech and music~\cite{diffwave21,SpecDiff22}.
Even in symbolic music generation, Mittal et al.~\shortcite{DBLP:Mittal21} use a DDPM to model the continuous latent space of a symbolic music variational autoencoder.


Diffusion models have initially been proposed in both continuous and discrete formulations; however, discrete DPMs remain somewhat in the shadow of the great success of their continuous siblings. 

Hoogeboom et al.~\shortcite{hoog21} extended discrete DPMs from the binomial to the categorical domain and Austin et al.~\shortcite{DBLP:Austin21} proposed a more general framework for discrete  denoising diffusion probabilistic models (D3PM), which includes all previous definitions of D3PMs.
D3PMs are no mere theoretical innovation either; Taylor et al.~\shortcite{DBLP:Taylor22} demonstrate their capabilities on discrete sequences.
These models, and especially the absorbing state-based formulation by Austin et al. \shortcite{DBLP:Austin21} provide the foundation for our work.


Building on these lines of research, we propose SCHmUBERT ( = Symbolic Creative Hierarchic music Unmasking Bidirectional Encoder Representation Transformer), a discrete denoising diffusion probabilistic model applied directly to the discrete domain of Symbolic Music.
To summarize, our contributions include:

\begin{itemize}
    \item The first application of Discrete Denoising Diffusion Probabilistic Models
    to the discrete domain of symbolic music generation.
    
    \item Non-autoregressive note-level modeling of symbolic music that allows for flexible applications including inpainting at note level and accompaniment generation:
    SCHmUBERT can infill arbitrary masked tokens from single notes or short motifs to accompaniment generation by masking entire tracks.
    
    \item State-of-the-art performance:
    Although our model has five times fewer trainable parameters than Mittal et al.'s \shortcite{DBLP:Mittal21}, both its unconditional generation results and its conditioned infilling results outperform those of their continuous reference model.
\end{itemize}

While our model performs very well in our quantitative evaluation, we want to caution against uncritical uptake. 
Employing a small algorithmic generation experiment, we discuss the limits of currently used evaluative metrics of symbolic music generation, especially concerning undue aggregation of values. 
We use an arbitrarily chosen starting distribution of notes and simulated annealing to create a piece that metrics fail to distinguish from a reference piece but is clearly visually and sonically different.

Highlighting a hard-to-quantify aspect of music generation---the model's capacity for interactivity and control---we present a small post-hoc conditioning experiment.
SCHmUBERT is susceptible to classifier guidance by separately trained classifiers. 
As a proof of concept, we use a note density classifier to successfully guide SCHmUBERT to predefined note densities per measure with moderate loss of sample quality.

The remainder of this paper is structured as follows:
Section~\ref{sec:background} introduces symbolic music generation and discrete denoising diffusion probabilistic models.
Section~\ref{sec:experiment} details our model, representation, data, architecture, and training procedure.
The trained model is then compared to a baseline model \cite{DBLP:Mittal21} on two tasks - infilling and unconditional generation - the results of which are presented in Section~\ref{sec:results}.
Finally Section~\ref{sec:discussion} discusses the evaluation metrics in the broader context of symbolic music generation and concludes the paper.

\section{Background \& Related Work}\label{sec:background}

Computation was barely a theoretical vision when researchers started to muse about the possibility of composing music automatically and algorithmically, and the interest hasn't faded since (Ada Lovelace as cited in 
\cite{aiello2016multifaceted}).
Many algorithms for symbolic music generation have been investigated and published.
In recent years, most algorithmic systems have been based on AI techniques, either evolutionary algorithms or, more commonly, deep neural networks~\cite{civit2022systematic}.

\subsection{Symbolic Music Generation}

This section introduces several deep learning (DL) symbolic music generation models.
While we do not directly compare all these models, we want to give a minimal overview of the most influential DL models of symbolic music as they happen 
to roughly map the space of available representations, architectures, learning models, and generation paradigms:

MusicVAE \cite{MVAE18} is a music generator based on variational autoencoders (VAE).
VAEs \shortcite{Kingma2014} are latent space models consiting of an encoder and a decoder network.

MusicVAE uses bidirectional recurrent neural networks as encoders and decoders.
The model employs one of the earliest fixed-grid tokenized MIDI representations and produces musical material in two bar segments.


MidiNet takes a different approach and uses a generative adversarial network (GAN) \cite{MIDINET17,GAN14}.
Architecturally, MidiNet employs convolutional neural networks (CNN) for both generator and discriminator, and consequentially,
represents its symbolic music data as fixed-size piano rolls.


Attention-based Transformer encoders are the foundational architecture of several recently published non-autoregressive music generation models.
These models are sometimes presented as general pre-trained music understanding models providing embeddings suitable to diverse downstream tasks beyond music generation.
This is warranted by applying a key technique that enabled the success of large-scale language pre-training: non-sequential unmasking \cite{BERT19}.

MusicBERT is a Bidirectional Encoder Representation Transformer (BERT) that learns to predict masked tokens in a symbolic music sequence~\cite{zeng2021musicbert}
An octuple token is introduced that contains higher-level information than just the note (e.g. time signature and tempo), the masking occurs at measure level. 

MidiBERT-Piano is another BERT-style unmasking model~\cite{chou2021midibert}. 
MidiBERT-Piano is trained exclusively on piano music - single track and non-monophonic - and evaluated both on generation and downstream classification tasks.


Mittal et al. \shortcite{DBLP:Mittal21} proposed a symbolic music generation system using DDPMs.
This model builds on the two-bar monophonic MusicVAE proposed earlier \cite{MVAE18}.
A continuous DDPM is then trained on sequences of latent MusicVAE embeddings.
This model was designed for longer-term structure via hierarchical modeling --- steering the generation process of a latent space model (MusicVAE) with another model--- and uses DDPMs in the continuous (latent) domain.

To the best of our knowledge, Mittal et al.'s is the only published diffusion model for symbolic music generation, albeit with the diffusion model applied indirectly.
It serves as our reference model: We keep a MusicVAE-like token representation and Mittal et al.'s evaluation metrics \shortcite{DBLP:Mittal21} .

\subsection{Diffusion Probabilistic Models}

Diffusion Probabilistic Models (DPMs) generate data by inverting a Markovian data corruption process.
Although Sohl-Dickstein et al.~\shortcite{DBLP:SD15} introduced DPMs for continuous and discrete (binomial) domains, the former---under the name of Denoising Diffusion Probabilistic Models (DDPMs)---have received relatively more attention in recent work \cite{DBLP:Ho20,DBLP:Mittal21,DBLP:Dhariwal21}.
Hoogeboom et al.~\shortcite{hoog21} extended DDPMs from the binomial to the categorical domain, and Austin et al.~\shortcite{DBLP:Austin21} propose a more general framework for Discrete DDPMs (D3PMs), which includes all previous definitions of D3PMs.

The Markovian data corruption process $q(\textbf{x}_t \vert \textbf{x}_{t-1})$ is called forward diffusion process, and in $T$ steps transforms data from the complex target distribution $p(\textbf{x}_0)$ into a distribution $p(\textbf{x}_T)$, from which sampling without access to target data is possible.
Once the Markovian reverse diffusion process $p_\theta(\textbf{x}_{t-1}\vert\textbf{x}_t)$ is learned, it can be used to transform samples from $p(\textbf{x}_T)$ to samples from the target distribution $p(\textbf{x}_0)$.

\begin{figure}[h]
    \centering
    \includegraphics[scale=0.1]{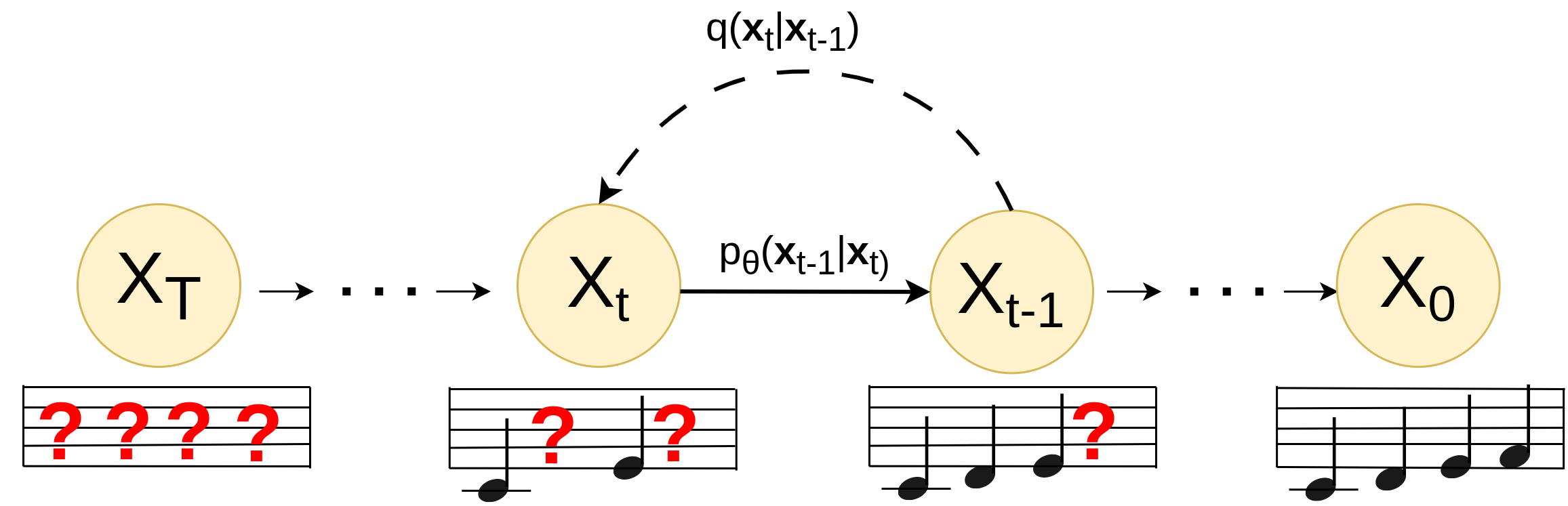}
    \caption{A score example illustrating the absorbing state diffusion process}
    \label{fig:for_rev_diff}
\end{figure}

In the continuous case, the forward diffusion process adds small amounts of Gaussian noise each step, resulting in $\textbf{x}_T$ converging towards an isotropic Gaussian distribution.
In \cite{DBLP:Austin21}'s framework for Discrete Denoising Diffusion Probabilistic Models, the forward transition probabilities for scalar discrete random variables $x_t \in \{1, ..., K\}$  are expressed using matrices: $q(x_t=j \vert x_{t-1}=i) = [\textbf{Q}_t]_{ij}$.
Given a set of random variables $\textbf{x}_t$ in one-hot notation, the forward diffusion process can be written as:

\begin{align}
    q(\textbf{x}_t \vert \textbf{x}_{t-1}) = \text{Cat}(\textbf{x}_t; \textbf{p} = \textbf{x}_{t-1}\textbf{Q}_t)
\end{align}
where $\text{Cat}(\textbf{x}, \textbf{p})$ is a categorical distribution, over the one-hot vector \textbf{x}, and probabilities are given by the vector $\textbf{p}$.
$\textbf{x}_{t-1}\textbf{Q}_t$ denotes the (one-hot) vector-matrix product and amounts to selecting a specific column from the transition matrix.
This transition matrix allows encoding domain-specific knowledge into the forward, and thus into the trainable reverse diffusion process: Transition probabilities between states reflect similarities between them.
\cite{DBLP:Austin21} define four flavors of D3PM, each characterized by its type of transition matrix:

The \textbf{uniform D3PM} has a matrix with uniform transition probabilities between all categories, $p(\textbf{x}_T$ is a uniform distribution over all categories.

The \textbf{discretized Gaussian} D3PM samples transition probabilities from a Gaussian PDF, and thus imposes higher transition probabilities between closer states.
This is very suitable for ordinal data, such as brightness in image pixels.\\
The \textbf{Absorbing state} D3PM (see figure \ref{fig:for_rev_diff}) introduces an additional category that does not occur in the domain: the Absorbing state.
This Absorbing state indicates corrupted data, and is similar to the [MASK] token in BERT-like models \cite{BERT19}.
In the forward diffusion process, variables are masked out randomly, they can only transition into the Absorbing state, but not leave it (thus absorbing).
In the reverse diffusion process, on the other hand, variables can only transition if they are in the Absorbing state.\\
\textbf{Token embedding distance} D3PMs are suitable where similarities between categories are not ordinal, but still can be distinguished.
Austin et al. mention text at character level encoding,  where vowels are likely to be more similar to each other than consonants as an example.
These similarities can be introduced to the forward transition matrices.

\subsubsection{Training}
The generative distribution $p_\theta(\textbf{x}_{t-1}\vert\textbf{x}_t)$ can be approximated using a neural network.
Given an $\textbf{x}_t$, the neural network predicts the reverse diffusion transition probabilities.\\

With access to $\textbf{x}_0$, these reverse transition probabilities can be explicitly expressed using the forward diffusion process:

\begin{align}
    q(\textbf{x}_{t-1} \vert \textbf{x}_t, \textbf{x}_0) &= \frac{q(\textbf{x}_{t} \vert \textbf{x}_{t-1}, \textbf{x}_0) q(\textbf{x}_{t-1} \vert \textbf{x}_0)}{q(\textbf{x}_t \vert \textbf{x}_0)} \\&= \text{Cat}\Big( \textbf{x}_{t-1}; \textbf{p} = \frac{\textbf{x}_t \textbf{Q}_t^\top \circ \textbf{x}_0 \bar{\textbf{Q}}_{t-1}}{\textbf{x}_0 \bar{\textbf{Q}}_{t} \textbf{x}_t^\top}\Big)
\end{align}
As the forward diffusion process is defined as a Markov chain, $q(\textbf{x}_{t-1} \vert \textbf{x}_t, \textbf{x}_0) = q(\textbf{x}_{t-1} \vert \textbf{x}_t)$.

Training the generative distribution amounts to minimizing the negative log-likelihood of real samples, which typically is done indirectly by minimizing the evidence lower bound (ELBO):
\begin{align}
L_{\mathrm{vb}}=&\mathbb{E}_{q\left(\boldsymbol{x}_0\right)}[\underbrace{D_{\mathrm{KL}}\left[q\left(\boldsymbol{x}_T \mid \boldsymbol{x}_0\right) \| p\left(\boldsymbol{x}_T\right)\right]}_{L_T}+\nonumber \\
&\sum_{t=2}^T \underbrace{\mathbb{E}_{q\left(\boldsymbol{x}_t \mid \boldsymbol{x}_0\right)}\left[D_{\mathrm{KL}}\left[q\left(\boldsymbol{x}_{t-1} \mid \boldsymbol{x}_t, \boldsymbol{x}_0\right) \| p_\theta\left(\boldsymbol{x}_{t-1} \mid \boldsymbol{x}_t\right)\right]\right]}_{L_{t-1}} \nonumber \\
& \underbrace{-\mathbb{E}_{q\left(\boldsymbol{x}_1 \mid \boldsymbol{x}_0\right)}\left[\log p_\theta\left(\boldsymbol{x}_0 \mid \boldsymbol{x}_1\right)\right]}_{L_0}]
\end{align}
A typical DDPM training process minimizes $L_{\mathrm{vb}}$ stochastically by minimizing randomly sampled subterms.
For efficient sampling of suberterms of $L_{\mathrm{vb}}$, $x_t$ can be inferred from $x_0$ in closed form in a single step:

\begin{align}
    q(\textbf{x}_t \vert \textbf{x}_0) = \text{Cat}(\textbf{x}_t; \textbf{p} = \textbf{x}_0 \bar{\textbf{Q}}_t) & & \text{,where } \bar{\textbf{Q}}_t = \prod_{i=1}^t \textbf{Q}_i
\end{align}

\subsubsection{Connection to other generative models}
Although the sampling process of D3PMs is very similar to that of autoregressive models, there is a crucial difference:
Autoregressive models are usually optimized to predict the next token in a sequence, while DDPMs refine their whole, joint output sequence simultaneously.
This is a consequence of how their tasks are formulated: DDPMs are optimized for fixed sequence lengths. In contrast, the sequence length in autoregressive models is flexible both during training and inference.
An Absorbing state D3PM could be used to sample autoregressively, by unmasking tokens in a fixed (linear) order at the cost of sample quality and diversity \cite{DBLP:Austin21}.

Diffusion models share some striking similarities with VAEs. 
Loosely speaking, diffusion models can be seen as Variational Autoencoders with their latent dimension equal to their data dimension.
Their encoder is $T$ layers deep and has no trainable parameters, while their decoder holds all trainable parameters and has a depth of $T$ times the depth of $p(\textbf{x}_{t-1}\vert\textbf{x}_t)$ ($T$ repeated applications).
 The model $p(\textbf{x}_{t-1}\vert\textbf{x}_t)$ is trained at each of the $T$ layers using intermediate results from the forward diffusion process, which avoids the vanishing gradient problem of neural networks with such large depths.
 At the other extreme of depth are one-step diffusion models.
 In particular, BERT \cite{BERT19} can be seen as a one-step Absorbing state diffusion model: its Absorbing state is the [MASK] token, which the model fills in in a single step \cite{DBLP:Austin21}.

\section{Experiments}\label{sec:experiment}

In this section, we introduce the components of SCHmUBERT.
We discuss the D3PM employed, the data representation, model architecture, and training hyperparameters.
In the section \ref{sec:results}, we compare two models against a state-of-the-art competitor~\shortcite{DBLP:Mittal21} in both unconditional generation and conditional infilling tasks.

\subsection{Absorbing State Diffusion Probabilistic Model}
We choose our model to be an Absorbing State Denoising Diffusion Probabilistic Model (ASD3PM), as this offers three advantages over the other flavors of D3PMs:
Firstly, ASD3PMs are capable of directly infilling in any sample containing masks.
Secondly, ASD3PMs can be trained using a simplified loss parameterized only on $\textbf{x}_0$.
Thirdly, this simplified parametrization allows adapting the number of diffusion steps at sample time \cite{DBLP:Taylor22}.

In our ASD3PM, the forward diffusion process $q(\textbf{x}_t \vert \textbf{x}_{t-1})$ masks out tokens with probability $\frac{t}{T}$, while the reverse diffusion process $p_\theta(\textbf{x}_{t-1}\vert\textbf{x}_t)$ recovers the masked out tokens.
Like Bond-Taylor et al.~\shortcite{DBLP:Taylor22}, we directly predict $p_\theta(\textbf{x}_0\vert\textbf{x}_t)$, which in training enables respecting the error for all masked out tokens, instead of only those unveiled in the current reverse diffusion step.
Prediction of all tokens jointly models all possible $\textbf{x}_{t-1}$ instead of only a specific one.
The single step prediction $p_\theta(\textbf{x}_{t-1}\vert \textbf{x}_t)$ can be sampled from $p_\theta(\textbf{x}_0\vert\textbf{x}_t)$ by keeping predicted tokens in $\textbf{x}_0$ only with probability $\frac{1}{T-t}$.
Bond-Taylor et al.~\shortcite{DBLP:Taylor22} further find that assigning less importance to the loss at timesteps close to $T$ (many masks) boosts training.
Hence, they reweigh the ELBO by $\frac{T-t-1}{T}$:

\begin{align}
\mathbb{E}_{q\left(\textbf{x}_0\right)}\left[\sum_{t=1}^T \frac{T-t-1}{T} \mathbb{E}_{q\left(\textbf{x}_t \mid \textbf{x}_0\right)}\left[\sum_{\left[\textbf{x}_t\right]_i=m} \log p_\theta\left(\left[\textbf{x}_0\right]_i \mid \textbf{x}_t\right)\right]\right]\label{eq:l_ce_d3pm}
\end{align}
The inner sum in equation \ref{eq:l_ce_d3pm} forms the joint (log) probability of all random variables in $\textbf{x}_0$ given $\textbf{x}_t$, which results in the cross-entropy between $\textbf{x}_0$ and $\textbf{x}_t$ under the expectation of $\mathbb{E}_{q\left(\textbf{x}_t \mid \textbf{x}_0\right)}$.

\subsection{Data \& Representation}

We train our models on the Lakh MIDI dataset (LMD) \cite{LMD16}, which consists of more than 170,000 multi-track MIDI files, and to our best knowledge, remains the largest publicly available symbolic music dataset.
Following our reference model \cite{DBLP:Mittal21,MVAE18}, we use a time-quantized representation with 16 steps per bar resolution.
We extract $\frac{4}{4}$ monophonic melodies and trios with lengths between 16 and 64 bars using Magenta's MusicVAE \cite{MVAE18} pipelines.
The trios consist of a monophonic melody, a monophonic bass track, and a polyphonic drum track.
Each step of the melody and bass track can have one of 90 values: 88 pitch values, 1 value for note-off, and 1 to continue the previous note.
The drum track encodes note-on for 9 independent drums simultaneously and hence can have one of 512 ($=2^9$) values per step.
In sum, a melody piece with a length of 64 bars is a sequence of 1024 indices, while a trio piece of the same length is a sequence of 1024 triplets of indices.
We randomly transpose the melody and bass tracks to augment our training data.

\subsection{Model Architecture}

\begin{figure}[h]
    \centering
    \includegraphics[scale=0.094]{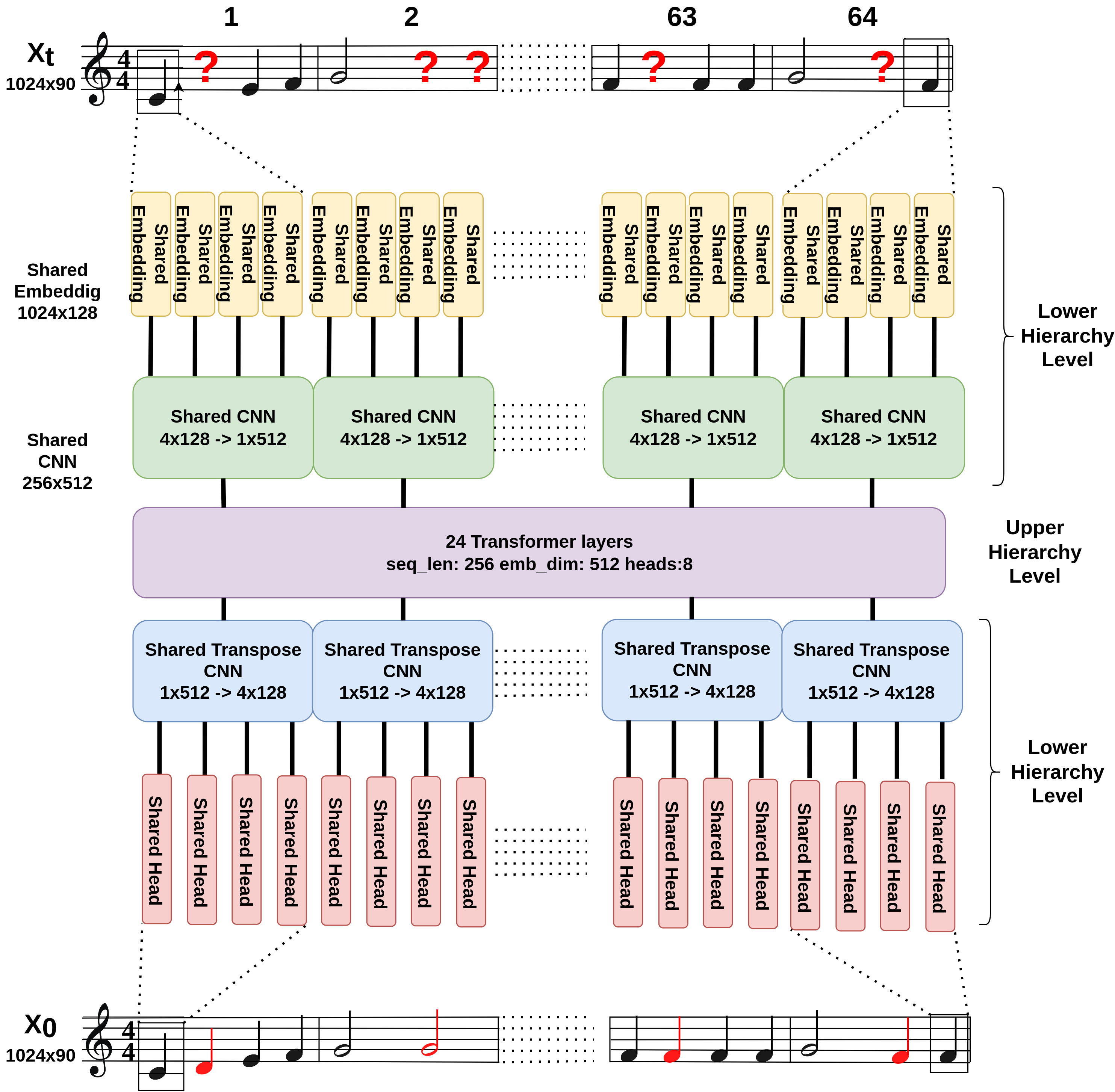}
    \caption{SCHmUBERT architecture. Four adjacent embeddings are convolved into one representation whether they encode a quarter note (such as the examples in the figure) or any other musical material.}
    \label{fig:SCHmUBERT}
\end{figure}

Figure \ref{fig:SCHmUBERT} shows the melody-only architecture of the neural network estimating transition probabilities of $p(\textbf{x}_0\vert\textbf{x}_t)$.
First, indices are embedded in a 128-dimensional vector.
A 1D CNN with kernel size 4 and stride 4 then computes first summarization---which we refer to as the lower hierarchy level--- of 4 adjacent 128-dimensional embeddings into one 512 dimensional embedding, effectively reducing the sequence length to a quarter, while increasing the embedding dimension.
The upper hierarchy level, a GPT-like \cite{GPT3} stack of 24 transformers processes the sequence of compressed embeddings.
The shared 1D Transpose CNN decompresses the sequence to its original length before a shared head maps the sequence of 1024x128 back to a sequence of indices.

In the trio case, each track passes through its own lower hierarchy level; 3 separate embeddings and 1D CNNs compress the track sequences independently.
The compressed outputs are then summed up before the upper hierarchy level processes them.
Three separate 1D Transpose CNNs compute the decompression, and separate heads predict the output probabilities.

\subsection{Training}
We train two versions of our model: a melody-only, and a trio version.
To allow for direct comparison with Mittal et al.'s model~\shortcite{DBLP:Mittal21}, the melody-only version is trained and evaluated on an extract of $1,000,000$ melodies without augmentation.\\
In table \ref{results:training:melody}, we provide further parameters of both training runs.

\begin{table}[h]
\scalebox{0.72}{
\begin{tabular}{ |c |c |c|}
\hline
Name & Melody & Trio\\
\hline
Melody/Bass pitches & 90&90\\
\hline
Augmentation & no & yes\\
\hline
Piece shape & (1024, 1) & (1024, 3)\\
\hline
Diffusion Model timesteps & 1024 & 1024\\
\hline
Optimizer & Adam @ lr$=5*10^{-4}$ & Adam @ lr$=5*10^{-4}$\\
\hline
Transformer layers & 24 & 24\\
\hline
Transformer embedding size & 512 & 512\\
\hline
Transformer attention heads & 8 & 8\\
\hline
Total parameters & 76,601,088 & 78,303,488\\
\hline
Train steps & 190,000 & 600,000\\
\hline
Validation loss & 0.087 & 0.046\\
\hline
Dataset size & 19000 Batches & 37649 Batches\\
\hline
Batch size & 50 & 50\\
\hline
GPU & 4x NVIDIA 2080 Ti & 4x NVIDIA 2080 Ti\\
\hline
Duration & $24$h & $50$h\\
\hline
\end{tabular}}
\caption{Parameters of training runs}\label{results:training:melody}
\end{table}

\section{Evaluation}\label{sec:results}

In this section, we compare our two models---melody-only and trio---against a state-of-the-art competitor~\shortcite{DBLP:Mittal21} in both unconditional generation and conditional infilling tasks.

\subsection{Metrics}

We evaluate our models using Mittal et al.'s~\shortcite{DBLP:Mittal21} framewise self-similarity metrics.
This metric captures local self-similarity patterns and is based on statistics of Overlapping Areas of Gaussian density functions.
We first define a sequence of 4-bar windows with a hop size of 2 bars.
For each window $w$, we model pitch and duration with a Gaussian distribution, respectively: $\mathcal{N}(\mu_{P_w}, \sigma^2_{P_w}), \mathcal{N}(\mu_{D_w}, \sigma^2_{D_w})$.
The Overlapping Area of Gaussian density functions of adjacent windows is then calculated.
The mean $\mu_{OA}$ and variance $\sigma^2_{OA}$ of overlap areas are then aggregated over all window tuples of a set of musical pieces.
From these values, the relative normalized distances \textit{Consistency} and \textit{Variance} are calculated as follows:

\begin{align}
    \textit{Consistency} &= \max(0, 1 - \frac{|\mu_{OA} - \mu_{GT}|}{\mu_{GT}})\\
    \textit{Variance} &= \max(0, 1 - \frac{|\sigma^2_{OA} - \sigma^2_{GT}|}{\sigma^2_{GT}})
\end{align}
where the values $\mu_{GT}$ and $\sigma^2_{GT}$ refer to the mean and variance of overlap areas in the ground truth data, respectively.
Overall, we compute four values per experiment: Consistency and Variance for pitch distributions and Consistency and Variance for duration distributions.

\subsection{Results}

\begin{figure}[h]
    \centering
    \includegraphics[scale=0.12]{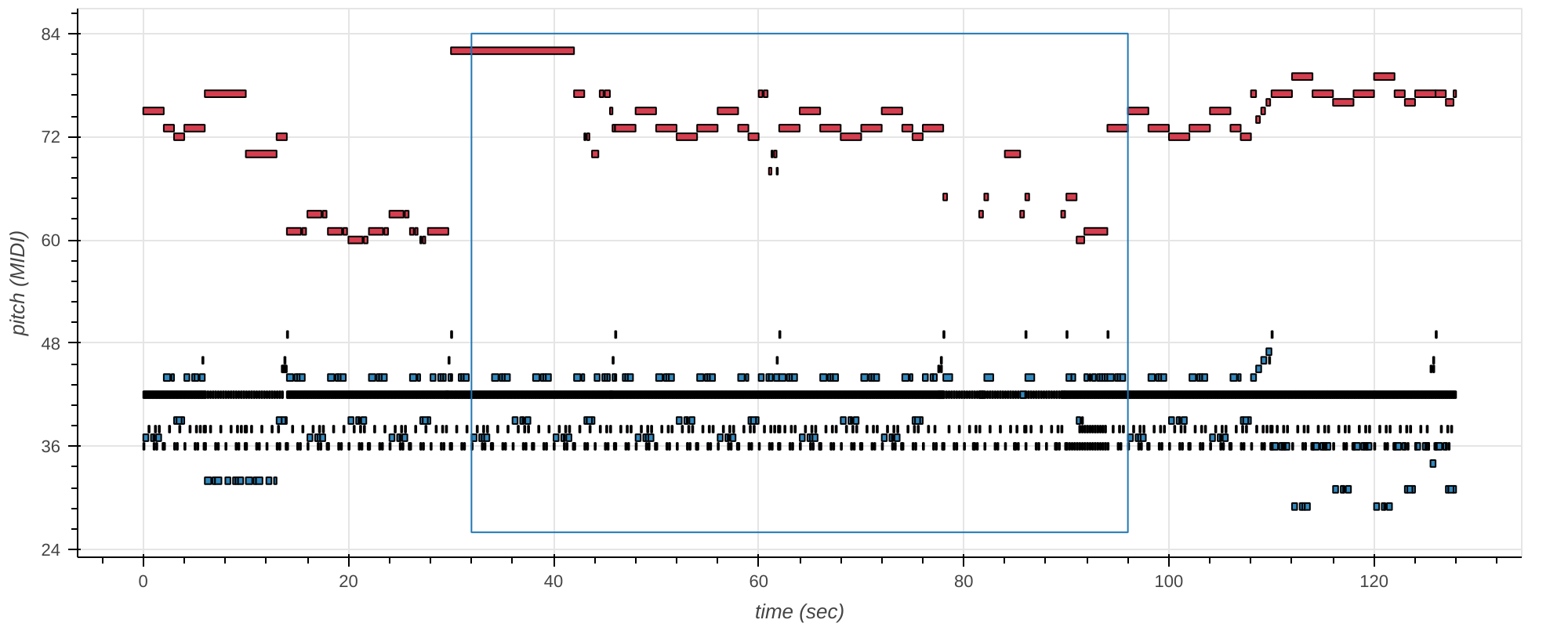}
    \includegraphics[scale=0.12]{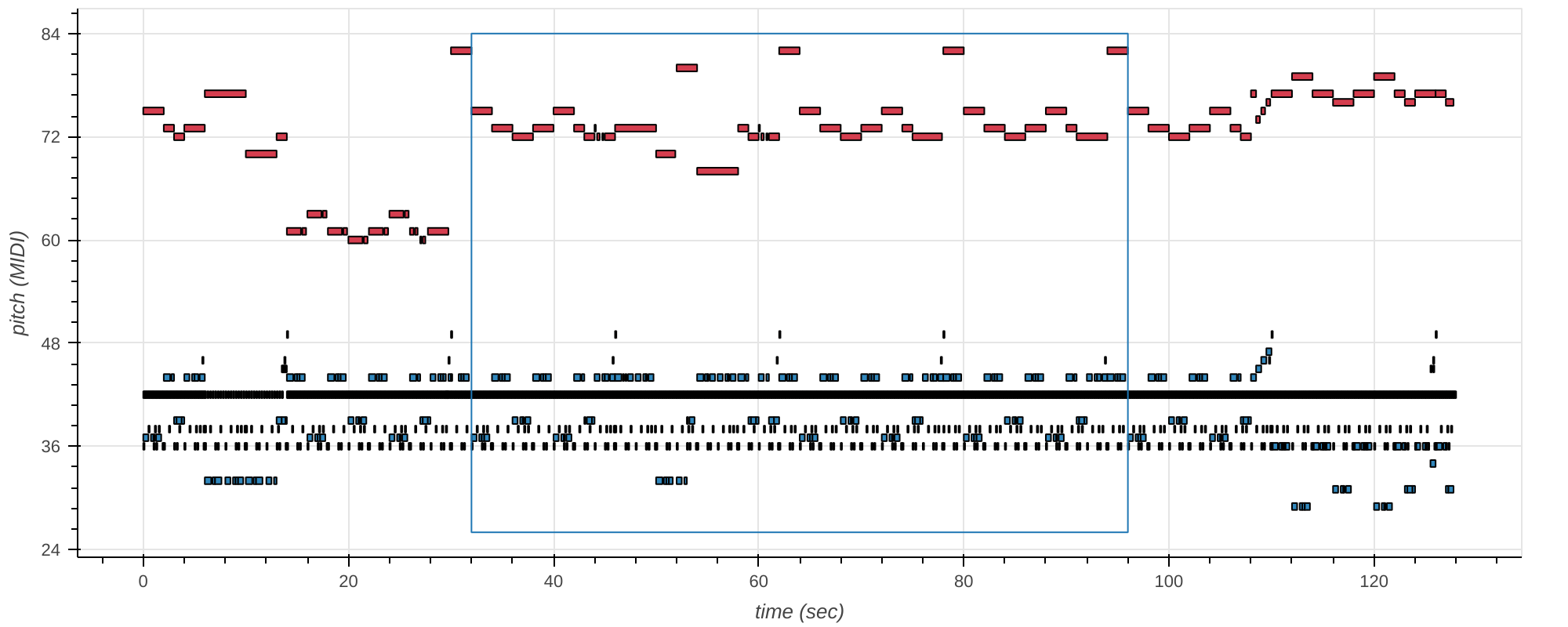}
    \caption{Trio infilling scenario: The top plot shows an original validation sample.
    The central 512 tokens (blue box) were masked out in all tracks, and filled in again by SCHmUBERT (bottom).
    The tracks are color-coded (red=melody, blue=bass, black=drum).
    Bass and drum tracks lie in similar pitch regions of the piano roll due to standard MIDI drum encoding.}
    \label{fig:trio_infilling}
\end{figure}

Like in \cite{DBLP:Mittal21}, we evaluate the self-similarity metric for sets of $1000$ generated pieces against $1000$ pieces drawn randomly from the joined train and evaluation set as ground truth.
In the infilling setting (Figure \ref{fig:trio_infilling}), the central 512 notes are masked and filled in by the model.
Consistency and Variance are inferred using the original, unmasked samples as ground truth.

\begin{table}[h]
\scalebox{0.6}{
    \begin{tabular}{ |c|c|c|c|c|c|c|c|c|  }
 \hline
 \rowcolor{gray!25}Setting & \multicolumn{4}{|c|}{Unconditional} & \multicolumn{4}{|c|}{Infilling} \\
 \hline
 \rowcolor{gray!25}Quantity & \multicolumn{2}{|c|}{Pitch} & \multicolumn{2}{|c|}{Duration} & \multicolumn{2}{|c|}{Pitch} & \multicolumn{2}{|c|}{Duration} \\
 \hline
 \rowcolor{gray!25}Metric & C & Var & C & Var & C & Var & C & Var \\
 \hline
 Train Data & 1.00 & 1.00 & 1.00 & 1.00 & 1.00 & 1.00 & 1.00 & 1.00 \\
 Test Data & 1.00 & 0.96 & 1.00 & 0.96 & 1.00 & 0.96 & 1.00 & 0.91 \\
  \hline
 \rowcolor{gray!25}Melody 64 bar & \multicolumn{8}{|c|}{}\\
 \hline
 Melody MDMVAE 64 bar & 0.99 & 0.90 & 0.96 & 0.92 & 0.97 & 0.87 & 0.97 & 0.80 \\
 \hline
\textbf{SCHmUBERT} & \textbf{0.992} & \textbf{0.920} & \textbf{0.993} & \textbf{0.937} & \textbf{0.997} & \textbf{0.970} & \textbf{0.996} & \textbf{0.970} \\
 \hline
 \rowcolor{gray!25}Trio 64 bar &  \multicolumn{8}{|c|}{}\\
\hline
\textbf{SCHmUBERT} & \textbf{0.996} & \textbf{0.893} & \textbf{0.990} & \textbf{0.896} & \textbf{0.997} & \textbf{0.964} & \textbf{0.995} & \textbf{0.924} \\
 \hline
 \textbf{Melody infilling} & - & - & - & - & \textbf{0.999} & \textbf{0.999} & \textbf{0.978} & \textbf{0.973} \\
 \hline
 \textbf{Bass infilling} & - & - & - & - & \textbf{0.999} & \textbf{0.999} & \textbf{0.991} & \textbf{0.996} \\
 \hline
  \textbf{Drum infilling} & - & - & - & - & \textbf{0.999} & \textbf{0.999} & \textbf{0.991} & \textbf{0.996} \\
 \hline
\end{tabular}}
\caption{\label{tab:results:eval} Self similarity scores, values in \textbf{bold} represent models proposed in this work}\label{tab:self_sim}
\end{table}

Table \ref{tab:self_sim} shows that SCHmUBERT consistently outperforms Mittal et al.'s Diffusion on MusicVAE-latents model (marked Melody MDMVAE) in the framewise self-similarity metric.
This is particularly remarkable when comparing their number of trainable parameters: SCHmUBERT has roughly 80 MIO trainable parameters, while \cite{DBLP:Mittal21}'s hybrid model has roughly 425 MIO trainable parameters (25MIO in the continuous DDPM, 400MIO in MusicVAE).
Inpainting and accompaniment experiments show that SCHmUBERT consistently deals with various scenarios.

We find that lower hierarchy level summarization significantly boosts training efficiency, while not reducing the model's capacity compared to a model without the lower hierarchy level and an embedding dimension of 512.
In the Trio setting, we alter the forward diffusion process: instead of masking notes uniformly in all three tracks, notes are either masked in one, two, or three tracks.
This masking schedule improves performance in the accompaniment setting.

\section{Discussion}\label{sec:discussion}

In this paper, we introduce a novel diffusion model that directly models the discrete variables of symbolic music.
We show this model to outperform a reference model on two tasks.
However, in place of concluding this paper with a simple recapitulation of the qualities of our models, we want to shed light on some critical aspects of music generation evaluation.
In this section, we will focus on two specific questions, discussing each from a general perspective first, then demonstrating them with particular small experiments in the following subsections.
These experiments are not evaluated thoroughly, but only serve as proofs of concept.

Starting from a general view on limitations of \textit{statistical quality metrics} for music generation (Section \ref{sec:subsec1}), we design an algorithm to question the merits of our own evaluation metrics specifically (Section \ref{sec:subsec2}).
Thereafter, we briefly discuss the idea of \textit{model affordances}, i.e.~(speculative) possibilities for user interaction and control (Section \ref{sec:subsec3}), 
before presenting a proof of concept for post-hoc conditioning on information not previously seen by the model (Section \ref{sec:subsec4}).

\subsection{Limitations of Statistical Metrics}\label{sec:subsec1}

The metrics used in the last five years' most cited symbolic music generation papers are negative log-likelihood, a set of musical metrics (ratio of empty bars, pitch classes per bar, qualified notes ratio, ratio of notes in patterns, tonal distance), and reconstruction accuracy, among various small-scale ablation studies~\cite{civit2022systematic}.
Two papers note no quantitative evaluation and instead opted for a listening test.

Given the variety of tasks and uses for symbolic music generation, it is unsurprising that no consensus on evaluation exists.
However, the difficulties extend beyond mere disagreement and have been addressed variously in the literature \shortcite{Agres2016,Pearce2001a,Sturm2016c}.
This section discusses the inherent theoretical and technical limitations of statistical metrics.

Following Naem et al.~\shortcite{naeem2020reliable}, we conceive of the quality of generative models as relating to two high-level concepts: Fidelity and diversity.
Fidelity denotes sample quality.
For music generation, this might include attributes like structuredness, melody-harmony consistency, correctness, and richness. 
Listening tests commonly assess such attributes, e.g.~in \cite{hsiao2021compound}.

Diversity, on the other hand, indicates a distributional want: the model should be able to produce a large variety of samples.
In probabilistic terms, it shouldn't suffer from mode collapse.
A somewhat related aspiration is originality, an operative definition of which is that the model should not produce copies of training data samples.

The metrics commonly used in the literature for the evaluation of music generation seem to cover neither fidelity nor diversity satisfactorily.
Most metrics are aggregated statistical estimates, which seem unlikely to depict musical concepts at their most granular---what musician ever thought of mismatching pitch class distributions in adjacent measures?---, and only become more impenetrable when aggregated over subsequent windows or even sets of pieces.
Arguably, such broad-brush criticism is as easy to dismiss as the metrics themselves --- after all, what is a \textit{musical} metric? ---, but we want to give a simple example of a failure of a specific metric: 
the very framewise self-similarity metric we used to evaluate our models.

\subsection{Confounding Consistency and Variance Metrics}\label{sec:subsec2}

\begin{figure}[h]
    \centering
    \includegraphics[width=\columnwidth]{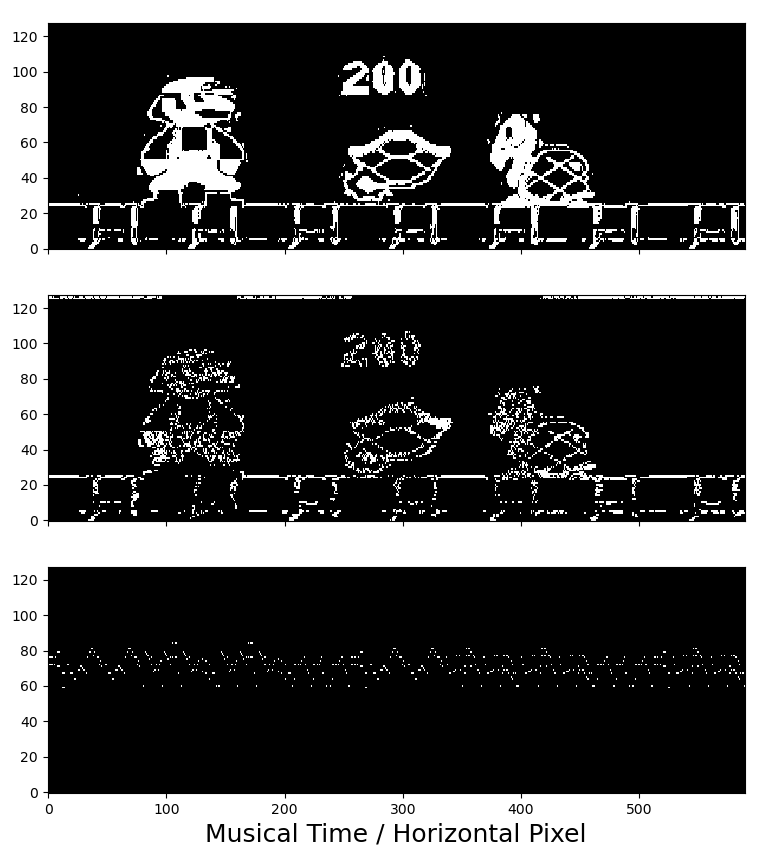}
    \caption{The Super Mario Bros video game (top), the Super Mario theme (bottom piano roll), and a sampled set of notes (middle piano roll) the self-similarity metrics cannot distinguish from the theme.}
    \label{fig:mario}
\end{figure}

In the following, we describe a simple algorithm that creates clearly spurious, yet high-ranking ``musical' material.
We take inspiration from Anscombe's famous quartet and its recent generalization based on simulated annealing~\cite{matejka2017same}.
Anscombe's quartet consists of four scatter plots of 2D datasets with similar descriptive statistics but visually different distributions.
Adapted to music generation, we devise a simulated annealing process to create a score with nearly indistinguishable consistency and variance values from those of a reference score in the dataset while visually similar to a chosen reference image.

The generation process starts with any binarized image in an appropriate pianoroll-like resolution (e.g. 128 X 500).
We compute "music" from the image by sampling column-wise from a pitch distribution given by a normalized and filtered vertical line of the image.
The resulting notes are then steered towards exhibiting similar Consistency and Variance metrics as a reference excerpt by repeatedly adding noise to pitches and durations and keeping the best proposals.

Figure \ref{fig:mario} illustrates the result of this process. 
Given an input image depicting a scene of a Super Mario Bros video game (top) and the Super Mario theme as a reference piece (bottom piano roll), this algorithm nudges a sampled set of notes (middle piano roll) toward being evaluatively indistinguishable from the reference piece while still visually close to the original image.
This shows a salient oversight of our quantitative evaluation metrics:
Surely we want them to be able to distinguish between Super Mario the image and Super Mario the theme.

With this result, we want to caution against expecting this metric to directly translate to "musical quality".
Note that due to the distributional nature of the metrics, this process is possible for a large variety of images and pieces.
We use this example to game the evaluation metrics used for our evaluation, however we conjecture that -- in keeping with Goodhart's law \cite{varela2014editorial} -- most distributional evaluation strategies are amenable to the same type of incursion.  
However, we cannot fully assess the impact of confoundable statistical metrics on
evaluations of generative models, including our own.

\subsection{Model Interaction and Control}\label{sec:subsec3}

In a second experiment, we discuss aspects of music generation quality that are harder to quantify but no less important; in fact, they are often pivotal for uptake by artists and instrument designers.
These aspects concern a generative system's potential for interaction and control.
The ASD3PM produces human-interpretable, partially masked note sequences in its intermediate steps, and thus affords flexible interactions not only before, but also during the sampling process. 
This is noteworthy, as such interactions are unobtainable in comparable generative models, which  rely on hard to interpret latent space representations \shortcite{MVAE18,DBLP:Mittal21}.

Our model allows for ex nihilo as well as accompaniment generation.
Furthermore, the sampling process can be carried out with unusually fine granularity, effectively enabling infilling as well as accompaniment generation with note-level resolution.
(\cite{DBLP:Mittal21}'s diffusion on MVAE-latents for example only supports infilling with a resolution of 32 notes at once.)
The generated sequence is of fixed length, but outpainting allows for length variability.

Diffusion models are inherently incremental, a suitable user interface can leverage this property and allow interaction with the sampling process at any timestep.
Consequently, these interactions can be adaptive, e.g. by iteratively resampling notes or motifs that a user has masked.

\subsection{Classifier Guidance}\label{sec:subsec4}

A particular advantage DPMs is flexible post-hoc generation conditioning.
One way of conditioning is guiding the generation process using gradients of a separately trained classifier.
For many DDPM formulations, this classifier needs to be trained on noisy data to provide correct gradients at different diffusion steps. 
however, the $\textbf{x}_0$ parametrization of our ASD3PMs allows using an off-the-shelf differentiable classifier that was only trained on real, unmasked data to guide the sampling process without the need for any retraining:

\begin{algorithm}[H]
    \caption{Guided Absorbing State D3PM Sampling}\label{alg:DMS}
    \textbf{Input:} $T$ steps, unmasking function $f_\theta(x_t)$\\differentiable loss function $l_\phi(x_0)$, guidance scale $s$
    \begin{algorithmic}
        \State $x_T \leftarrow [\text{MASK}]$
        \For{t = T, ..., 1}
            \State $x_{0 \text{logits}} = f_\theta(x_T)$
            \State $x_{0 \text{probs}} = \text{softmax}(x_{0 \text{logits}})$
            \State $loss = l_\phi(x_{0 \text{probs}})$
            \State $x_{0 \text{probs}} -= \nabla_{x0 \text{probs}} loss \cdot s$
            \State $x_{0 \text{cond}} = \text{Cat}(x_{0 \text{probs}}).\text{sample}()$
            \State with probability $\frac{1}{t} \text{replace masks in } x_T \text{with } x_{0 \text{cond}}$
        \EndFor
        \State \Return $x_T$
    \end{algorithmic}
\end{algorithm}

As a proof of concept, we train a small feed-forward neural network to predict note density, i.e.~the number of onsets per measure.
We then use this classifier to guide SCHmUBERT to predefined note densities.
Classifier guidance results in high accuracy, for plausible note densities (3-12 of max. 16), around 40\% of measures satisfy the target density, while almost all of the remaining measures count only one onset too many or little.
The subjective sampling quality apparently only decreases a little, but this needs further evaluation.

\subsection{Conclusion}

In this paper, we present SCHmUBERT, a discrete denoising diffusion probabilistic model applied directly to the modeling of symbolic music tokens.
SCHmUBERT outperforms reference work on two music generation tasks while being comparatively small and offering more options for conditioning.
Besides presenting a state-of-the-art model, we take a critical look at music generation evaluation and caution against lopsided evaluation based only on aggregated note statistics. 

\section*{Acknowledgments}

This work was supported by the European Research Council (ERC) under the EU’s Horizon 2020 research \& innovation programme, grant agreement No. 01019375 (\textit{Whither Music?}), and the Federal State of Upper Austria (LIT AI Lab).

\bibliographystyle{named}
\bibliography{ijcai23}

\end{document}